\documentclass[11pt]{article}

\usepackage[left=1.25in, right=1.25in, top=1in, bottom=1in]{geometry}
\usepackage{setspace,amsfonts,amsmath,graphicx,algorithm}
\renewcommand\footnotemark{}

\title{%
{\bf Distributed No-Regret Learning in\\  Multi-Agent Systems}\\
}

\author{\emph{Xiao Xu and Qing Zhao} \\ \emph{Cornell University, Ithaca, NY.} \emph{Email: \{xx243, qz16\}@cornell.edu}
\thanks{This work was supported by the National Science Foundation under Grant CCF-1815559.}}

\date{}

\begin{document}

\maketitle

\section{Introduction}

Game theory is a well-established tool for studying interactions among self-interested players.~Under the assumption of complete information on the game composition at each player, the focal point of game-theoretic studies has been on the \emph{Nash equilibrium}~(NE) in analyzing game outcomes and predicting strategic behaviors of rational players.  

The difficulty in obtaining complete information in real-world applications gives rise to the formulation of \emph{repeated unknown games}, where each player has access to only local information such as his own actions and utilities, but is otherwise unaware of the game composition or even the existence of opponents. In such a setting, a rational player improves his decision-making through real-time interactions with the system and learns from past experiences \cite{cesa2006prediction}. The problem can be viewed through the lens of distributed online learning, where the central question is whether learning dynamics of distributed players lead to a system-level equilibrium in some sense. Studies in the past few decades have revealed intriguing connections between various notions of \emph{no-regret} learning at each player and certain relaxed versions of NE at the system level \cite{cesa2006prediction, young2004strategic}.

While one-step closer to real-world systems, repeated unknown games, in their canonical forms, often adopt idealistic assumptions in terms of the stationarity of the player population and their utilities, availability of complete and perfect feedback, full rationality of players with unbounded cognition and computation capacity, and homogeneity among players in their knowledge of the game. Many emerging multi-agent systems, however, are inherently dynamic and heterogeneous, and inevitably limited in terms of available information and the cognition and computation capacity of the players. We give below two examples.

\vspace{.2cm}
\noindent\emph{Example: adversarial machine learning.} Security issues are at the forefront of machine learning and deep learning research, especially in safety-critical and risk-sensitive applications. The interaction between the defender and the attacker can be modeled as a two-player game. While the player population may be small, the game is highly complex in terms of the action space, utilities, feedback models, and the available knowledge each player has about the other. In particular, the attacker is characterized by its knowledge---how much information it has for designing attacks---and power---how often a successful attack can be launched. Both can be dynamically changing and adaptive to the strategies of the defender. A full spectrum of attacker profiles has been considered, ranging from the so-called black-box model to the white-box model (i.e., an omniscient attacker). The attack process is also dynamic, often exhibiting bursty behaviors following a successful intrusion or a system malfunction. The action space of the attacker can be equally diverse, including poisoning attacks and perturbation attacks. The former targets the training phase by injecting corrupted labels and examples for the purpose of embedding wrong decision rules into the machine learning algorithm. The latter targets the blind spots of a fully trained artificial intelligence using strategically perturbed instances that trigger wrong outputs, even when the perturbation is so minute as being indiscernible to humans. In terms of utilities, the attacker's goal may be to compromise the integrity of the system (i.e., to evade detection by causing false negatives) or the availability of the system (to flood the system with false positives).~See~a~comprehensive~taxonomy~of~attacks~against~machine~learning~systems~in~\cite{barreno2010security}.

\vspace{.2cm}

\noindent\emph{Example: transportation systems.} Route selection in urban transportation is a typical example of a non-cooperative game repeated over time. The game is characterized by a large population of players that is both dynamic and heterogeneous, with vehicles leaving and joining the system and utilities varying across players and over time. The envisioned large-scale adoption of autonomous vehicles will further diversify the traffic composition. Autonomous vehicles are significantly different from human drivers in terms of decision-making rationality, access to and usage of system-level knowledge, and memory and computation power. Bounded rationality is more evident in human drivers: they are likely to select a familiar route and inclined to settle for sufficing yet suboptimal options.

Complex multi-agent systems as in the above examples call for new game models, new concepts of regret, new design of distributed learning algorithms, and new techniques for analyzing game outcomes. We present in this article representative results on distributed no-regret learning in multi-agent systems. We start in Sec. \ref{sec:classic} with a brief review of background knowledge on classical repeated unknown games. In the subsequent four sections, we explore four game characteristics---dynamicity, incomplete and imperfect feedback, bounded rationality, and heterogeneity---that challenge the classical game models. For each characteristic, we illuminate its implications and ramifications in game modeling, notions of regret, feasible game outcomes, and the design and analysis of distributed learning algorithms. Limited by our understanding of this expansive research field and constrained by the page limit, the coverage is inevitably incomplete. We hope the article nevertheless provides an informative glimpse of the current landscape of this field and stimulates future research interests.

\section{Distributed Learning in Repeated Unknown Games}\label{sec:classic}

In this section, we review key concepts in game theory and highlight classical results on distributed learning in repeated unknown games.

\subsection{Static Games and Equilibria}

An $N$-player \emph{static game} is represented by a tuple $\mathcal{G}(\mathcal{N},\mathcal{A},u)$, where $\mathcal{N}=\{1,..., N\}$ is the set of players, $\mathcal{A}=\mathcal{A}_1\times\cdots\times\mathcal{A}_N$ the Cartesian product of each player's action space~$\mathcal{A}_i$, and $u=(u_1,...,u_N)$ the utility functions that capture the interaction among players. Specifically, the utility function $u_i$ of player $i$ encodes his preference towards an action. It is a mapping from the action profile ${\bf a}=(a_1,...,a_N)$ of all players to player $i$'s reward $u_i({\bf a})$. 

A \emph{Nash equilibrium} (NE) is an action profile ${\bf{a}}^*=(a^*_1,...,a^*_N)$ under which no player can increase his reward via a unilateral deviation. Specifically, $u_i({\bf a}^*)\ge u_i(a_i',{\bf a}^*_{-i})$ for all~$i$ and all $a_i'\neq a_i^*$, where ${\bf a}^*_{-i}$ denotes the action profile after excluding player $i$. Due to the focus on deterministic actions (also called \emph{pure strategies}), the resulting equilibrium is a \emph{pure Nash equilibrium}. A player may also adopt a \emph{mixed strategy}, which is a probability distribution $s_i$ over the action space. Correspondingly, a \emph{mixed Nash equilibrium} is a product distribution ${\bf{s}}^*=s^*_1\times\cdots\times s_N^*$ under which the expected utility $\mathbb{E}_{{\bf a}^*\sim{\bf s}^*}[u_i({\bf a}^*)]$ for every player $i$ is no smaller than that under a unilateral deviation $s_i'\neq s_i^*$ in player $i$'s strategy. A game with a finite population and a finite action space has at least one mixed NE but may not have any pure NE \cite{nisan2007algorithmic}.

NE is defined under the assumption that players adopt independent strategies (note the product form of ${\bf{s}}^*$). A more general equilibrium---\emph{correlated equilibrium} (CE)---allows correlation across players' strategies. We note that for equilibrium definitions introduced here, we focus on games with a finite action space. Specifically, a CE is a \emph{joint} probability distribution $\bf s$ (not necessarily in a product form) satisfying $\mathbb{E}_{{\bf a}\sim{\bf s}}[u_i(a_i, {\bf a}_{-i})|a_i]\ge\mathbb{E}_{{\bf a}\sim{\bf s}}[u_i(a_i',{\bf a}_{-i})|a_i]$ for all $i$, $a_i$, and $a_i'$, where the expectation is over the joint strategy $\bf{s}$ conditioned on that the realized action of player $i$ is $a_i$. The concept of CE can be interpreted by introducing a mediator, who draws an outcome $\bf a$ from $\bf s$ and privately recommends action $a_i$ to player~$i$. The equilibrium condition states that no player has the incentive to deviate from the outcome of the correlated draw from $\bf s$ after his part is revealed. CE can be further relaxed to the so-called \emph{coarse correlated equilibrium} (CCE), which is a joint distribution $\bf s$ satisfying $\mathbb{E}_{{\bf a}\sim{\bf s}}[u_i({\bf a})]\ge\mathbb{E}_{{\bf a}\sim{\bf s}}[u_i(a_i',{\bf a}_{-i})]$ for all $i$ and all $a_i'\neq a_i$. Different from CE, CCE imposes an equilibrium condition that is realization independent.

\begin{figure}
\begin{center}
\vspace{-.8cm}
\includegraphics[width = .75\textwidth]{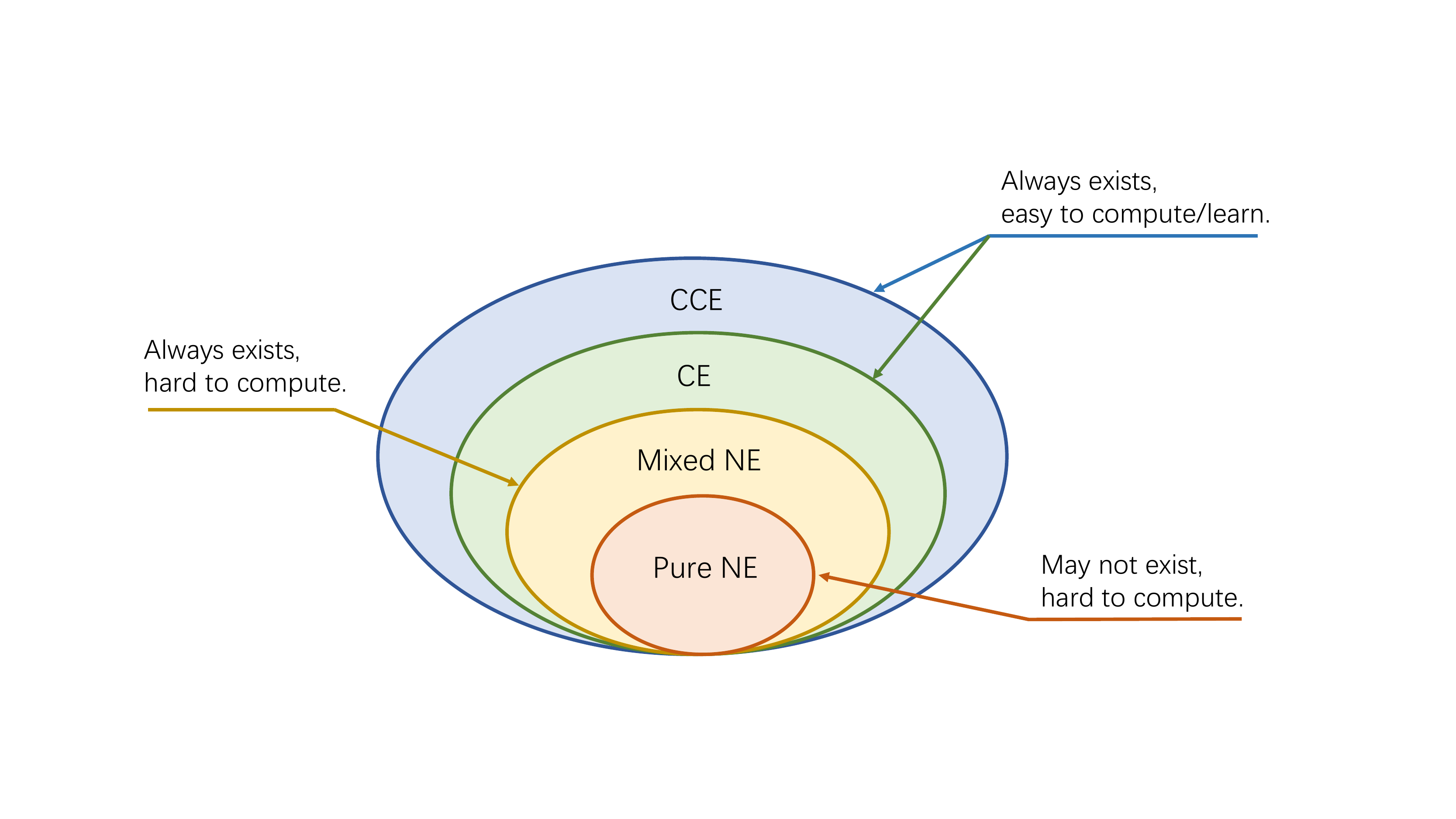}
\vspace{-1.5cm}
\caption{Relations and properties of four types of equilibria \cite{roughgarden2015intrinsic}.}\label{equilibria}
\end{center}
\vspace{-.5cm}
\end{figure}

The four types of equilibria exhibit a sequential inclusion relation as illustrated in Fig. \ref{equilibria}. The more general set of strategy profiles (i.e., allowing correlated strategies across players) in CE and CCE may lead to higher expected utilities summed over all players. CE and CCE can also be computed via linear programming, while pure NE and mixed NE are hard to compute \cite{nisan2007algorithmic}. More importantly, CE and CCE are learnable through certain learning dynamics of players when a game is played repeatedly as discussed next. A caveat is that the set of CCE may contain highly non-rational strategies that choose only strictly dominated actions (actions that are suboptimal responses to all action profiles of the other players). See \cite{viossat2013no} for specific examples.

\subsection{Repeated Unknown Games and No-Regret Learning}

A \emph{repeated game} consists of $T$ repetitions of a static game (referred to as the stage game in this context)\footnote{In a general definition of a repeated game \cite{laraki2015advances}, the stage game is parameterized by a state, which affects the utility function. Two basic settings exist in the literature: (i) the state evolves over time following a Markov transition rule (the state in the next stage depends on the state and actions in the current stage); (ii) the state is fixed throughout all stages. We focus on the second setting in discussing classical results on repeated games.}. In a repeated unknown game, after taking an action $a_i^t$ (potentially randomized according to a mixed strategy) in the $t$-th stage, player $i$ accrues a utility $u_i({\bf a}^t)$ and observes the entire utility vector $(u_i(a_i',{\bf a}^t_{-i}))_{a_i'\in\mathcal{A}_i}$ for all actions $a_i'$ in his action space (we focus on a finite action space here) against the action profile ${\bf a}_{-i}^t$ of the other players. The actions and utilities of the other players, however, are unknown and unobservable. 

From a single player's perspective, a repeated unknown game can be viewed as an online learning problem where the player chooses actions sequentially in time by learning from past experiences. A commonly adopted performance measure in online learning is \emph{regret}, defined as the cumulative reward loss against a properly defined benchmark policy with hindsight vision and/or certain clairvoyant knowledge about the game. In other words, the benchmark policy defines the learning objective that an online algorithm aims to achieve over time. Different benchmark policies lead to different regret measures. Two classical regret notions are the \emph{external regret} and the \emph{internal regret} as detailed below.

Let $\pi_i$ denote the online learning algorithm adopted by player $i$. For a fixed action sequence $\{{\bf a}_{-i}^t\}_{t=1}^{T}$ of the other players, the external regret of $\pi_i$ is defined as:
\begin{equation}
	\max_{a'\in\mathcal{A}_i}\mathbb{E}_{\pi_i}\left[ \sum_{t=1}^{T}(u_i(a',{\bf a}_{-i}^t)-u_i({\bf a}^t))\right],
\end{equation}
where $\mathbb{E}_{\pi_i}$ denotes the expectation over the random action process $\{a_i^t\}_{t=1}^{T}$ induced by $\pi_i$. In other words, the benchmark policy in the external regret chooses the best fixed response to the other players' actions in hindsight. The internal regret of $\pi_i$ is defined as:
\begin{equation}\label{Rin}
	\max_{a,a'\in\mathcal{A}_i}\mathbb{E}_{\pi_i}\left[\sum_{t=1}^{T}\mathbb{I}\{a_i^t=a\}(u_i(a',{\bf a}_{-i}^t)-u_i({\bf a}^t))\right],
\end{equation}
where $\mathbb{I}\{\cdot\}$ is the indicator function.~In~this~definition,~the~benchmark~policy~is~the~best~hindsight~\emph{modification}~of~$\pi_i$~by~swapping~a~single~action~with~another~throughout~all~stages.

An online learning algorithm $\pi_i$ is said to achieve the \emph{no-regret} condition if against all action sequences $\{{\bf a}_{-i}^t\}_{t=1}^{T}$ of the other players, the cumulative regret has a sublinear growth rate with the time horizon $T$. In other words, $\pi_i$ offers, asymptotically as $T\to\infty$, the same average reward per stage as the specific benchmark policy adopted in the corresponding regret measure. No-regret learning is also referred to as Hannan consistency due to the original work \cite{hannan1957approximation} as well as \cite{blackwell1956analog}. 

It is clear that the significance of no-regret learning depends on the adopted benchmark policy which the learning algorithm is measured against. A benchmark policy with stronger performance leads to a stronger notion of regret. In particular, the internal regret is a stronger notion than the external regret: no-regret learning under the former implies no-regret learning under the latter, but not vice versa \cite{stoltz2005internal}.

A number of no-regret learning algorithms exist in the literature. Representative algorithms achieving no-external-regret learning include \emph{Multiplicative Weights} (MW) (also known as the Hedge algorithm) and Follow the Perturbed Leader \cite{cesa2006prediction}. Both are randomized policies, as randomization is necessary for achieving no-regret learning in an adversarial setting with general reward functions \cite{cesa2006prediction}. In particular, under the MW algorithm, each player maintains a weight $W_a(t)$ of each action $a$ at every stage $t$ based on past rewards: $W_a(t)=e^{\epsilon \sum_{\tau=1}^{t}r_a(\tau)}=W_a(t-1)e^{\epsilon r_a(t)}$, where $r_a(\tau)$ is the reward received under $a$ at stage $\tau$ and $\epsilon>0$ is the learning rate. The probability of choosing $a$ in the next stage is proportional to its weight given by $\frac{W_a(t)}{\sum_{a'}W_{a'}(t)}$.

For no-internal-regret learning, a representative algorithm is \emph{Regret Matching} \cite{hart2000simple}. Let $R^{a\to a'}(t)=\frac{1}{t}\sum_{\tau=1}^{t}\mathbb{I}\{a_i^{\tau}=a\}(u_i(a',{\bf a}_{-i}^{\tau})-u_i({\bf a}^{\tau}))$ denote the average gain per play by switching from action $a$ to an alternative $a'$ in the past $t$ plays. In the ($t+1$)-th stage, the probability of switching from the previous action $a_t$ to an alternative $a'$ is given by $\frac{1}{\epsilon}R^{a_{t}\to a'}(t)$, where $\epsilon>0$ is a normalization parameter chosen to ensure a positive probability of staying with action $a_t$. Regret Matching also offers no-external-regret learning by setting the probability of selecting an action $a$ at the ($t+1$)-th stage to the normalized average gain per play from playing action $a$ throughout the past $t$ plays, i.e., $\frac{R^a(t)}{\sum_{a'}R^{a'}(t)}$,  where $R^a(t)=\frac{1}{t}\sum_{\tau=1}^{t}(u_i(a,{\bf a}_{-i}^{\tau})-u_i({\bf a}^{\tau}))$ \cite{hart2000simple}.

\subsection{System-Level Performance under No-Regret Learning}

Regret captures the learning objective of an individual player. At the system level, it is desirable to know whether the dynamical behaviors of distributed players converge to an equilibrium in some sense and whether the self-interested regret minimization promises a certain level of optimality in terms of social welfare.

For the first question, it has been shown that if every player adopts a no-external-regret learning algorithm, the empirical distribution of the sequence of actions taken by all players converges to the set of CCE of the stage game \cite{roughgarden2015intrinsic}.  No-regret learning under the internal regret measure guarantees convergence to the more restrictive set of CE \cite{hart2000simple}. Such convergence results are, however, in terms of the empirical frequency of the players' actions rather than the actual sequence of plays. The convergence is also only to the set of equilibria, rather than to an equilibrium in the corresponding set. In fact, by treating learning in games as a dynamical system, recent studies have shown that in the continuous-time setting, the actual plays under no-regret learning algorithms (such as Follow the Regularized Leader) may exhibit cycles rather than convergence \cite{mertikopoulos2018cycles}. In the discrete-time setting, it has been shown that in zero-sum games, the actual plays under the MW algorithm (starting from a non-equilibrium initial strategy) diverges from every fully mixed NE \cite{bailey2018multiplicative}. For games with special structures (e.g., potential games \cite{heliou2017learning} with a finite action space and bilinear smooth games \cite{gidel2019negative} with a continuum of actions), however, stronger results on the convergence of the actual plays to the more restrictive set of (mixed) NE have been~established.

In addition to the convergence of learning dynamics, the social welfare resulting from the self-interested learning of individual players is of great interest in many applications. In (known) static games, the loss in social welfare  $W({\bf s})=\mathbb{E}_{{\bf{a}}\sim{\bf s}}\left[\sum_{i=1}^{N}u_i({\bf a})\right]$ (i.e., the system-level utility under a strategy profile $\bf s$) due to the self-interested behaviors of players is quantified by the \emph{price of anarchy} (POA). It is defined as the ratio of the optimal social welfare $\textrm{OPT}=\max_{{\bf s}}W({\bf s})$ among all strategies to the smallest social welfare in the set of mixed NE. For repeated unknown games, a corresponding concept, \emph{price of total anarchy} (POTA), is defined as:
\begin{equation}\label{POTA}
\frac{\textrm{OPT}}{\min_{{\bf s}^1,...,{\bf s}^T}\frac{1}{T}\sum_{t=1}^{T}{W({\bf s}^t)}},
\end{equation}
where ${\bf s}^1,...,{\bf s}^T$ is the sequence of strategy profiles in the no-regret dynamics of all players. It has been shown that in games with special structures (e.g., valid games and congestion games), no-regret learning guarantees a POTA that converges to the POA of the stage game even though the sequence of actual plays may not converge to a (mixed) NE \cite{blum2008regret}. The convergence of the POTA to the POA of the stage game implies that no-regret learning can fully negate the impact of the unknown nature of the game on social welfare. The result was later extended in \cite{roughgarden2015intrinsic} to a general class of games referred to as \emph{smooth games} (which includes valid games and congestion games as special cases). To achieve higher social welfare, cooperation among players is necessary. For example, if every player agrees to follow a learning algorithm designed specifically for optimizing the system-level performance, the optimal action profile will be selected a high percentage of time \cite{marden2014achieving}.

\section{Dynamicity}\label{IS1}

In a dynamic repeated game, the stage game is time-varying. The dynamicity may be in any of the three elements of the game composition: the set of players, the action space, and the utility functions\footnote{Note that the general definition of repeated games in \cite{laraki2015advances} includes dynamicity in the utility function, as the state parameter may evolve over time following a Markov transition rule. The dynamic repeated game discussed in this section differs from the general repeated game in two aspects: (i) the set of players and the action space can also be time-varying; (ii) the utility functions are in general independent across stages.}.

\subsection{Notions of Regret}

Dynamic unknown games call for new notions of regret to provide meaningful performance measures for distributed online learning algorithms. Specifically, the benchmark policy of a fixed single best action used in the external regret and that of a fixed single best action modification used in the internal regret can be highly suboptimal in dynamic games. As a result, achieving no-regret learning under thus-defined regret measures can no longer serve as a stamp for good performance.

A rather immediate extension of the external regret is to consider every interval of the learning horizon and measure the cumulative loss against a single best action in hindsight that is specific to each interval. This leads to the notion of \emph{adaptive regret}, under which no-regret learning requires a sublinear growth of the cumulative reward loss in every interval as the interval length tends to infinity. The adaptive regret is particularly suitable for piecewise stationary systems where changes can be abrupt but infrequent. Classical learning algorithms such as MW can be extended to achieve no-adaptive-regret \cite{luo2015achieving}. The key issue in algorithm design is a mechanism to discount experiences from the distant past.

Another extension of the external regret is the so-called \emph{dynamic regret}, in which the benchmark policy can be an arbitrary sequence of actions, as opposed to a fixed action throughout an interval of growing length. Achieving diminishing reward loss against all sequences of actions is, however, unattainable. Constraints on either the benchmark action sequence or the reward functions are necessary for defining a meaningful measure. On the variation of the benchmark action sequence, a commonly adopted constraint in the setting with finite actions is that the benchmark sequence is piecewise-stationary with at most~$K$ changes (the thus-defined regret is also referred to as the \emph{K-shifting regret}). In this case, the no-adaptive-regret condition directly implies no-dynamic-regret \cite{luo2015achieving}. With a continuum of actions, the constraint is often imposed on the cumulative distance between every two consecutive actions in the sequence, i.e., $V_T(\{a^{t}\}_{t=1}^{T})=\sum_{t=1}^{T-1}||a^{t+1}-a^t||$. It has been shown that if the benchmark sequence is slow-varying, i.e., $V_T=o(T)$, no-dynamic-regret is achievable through well-designed restart procedures \cite{duvocelle2018learning}. The variation constraint can also be applied to the reward functions. A typical example with a continuum of actions is the sublinear ``variation budget'' assumption. Specifically, the cumulative variation between the reward functions in two consecutive stage games grows sublinearly in $T$, i.e., $\sum_{t=1}^{T-1}\sup_a|u_{t+1}(a)-u_{t}(a)|=o(T)$. Similar constraints can be imposed on the gradient $\nabla u_t(a)$ of the utility function and with the variation measured by the $L_p$-norm. See \cite{mokhtari2016online} and references therein for details and corresponding no-regret learning algorithms.

The external regret and its extensions are measured against an alternative strategy of a single player. A new notion of regret---\emph{Nash equilibrium regret}---considers  a benchmark policy that is jointly determined by the strategies of all players \cite{pmlr-v97-cardoso19a}. Consider a repeated game with time-varying utility functions $\{u_i^t\}_{t=1}^{T}$ for each player $i$. Let $\bar{u}_i=\frac{1}{T}\sum_{t=1}^{T}u_i^t$ be the average utility function and ${\bf s}^*$ the mixed NE of the static game defined by the average utility functions $\bar{u}=(\bar{u}_1,...,\bar{u}_N)$. The NE regret of player $i$ following a policy $\pi_i$ is then given by $\mathbb{E}_{\pi}[\sum_{t=1}^{T}u_i^t({\bf{a}}^t)]-T\mathbb{E}_{{\bf a}^*\sim{\bf s}^*}[\bar{u}_i({\bf{a}}^*)]$, where ${\bf a}^t$ is the action profile selected by the policies $\pi=(\pi_1,...,\pi_N)$ of all players at stage $t$. No-regret learning under the NE regret ensures that each player's average reward asymptotically matches that promised by the mixed NE under the average utility functions. A centralized learning algorithm achieving no-NE-regret was developed in \cite{pmlr-v97-cardoso19a} for repeated two-player zero-sum games with arbitrarily varying utility functions. Achieving no-regret learning under the measure of NE regret in a distributed setting, however, remains open.

\subsection{System-Level Performance}

The two key measures---convergence to equilibria and POTA---for system-level performance also need to be modified to take into account game dynamics. The time-varying sequence~$\{\mathcal{G}^t\}_{t=1}^{T}$ of stage games defines a sequence of equilibria and a sequence $\{\textrm{OPT}^t\}_{t=1}^{T}$ of optimal social welfare. The desired relation between no-regret learning dynamics at individual players and the system-level equilibria is thus in terms of tracking rather than converging. For the definition of POTA, the optimal social welfare in the numerator in (\ref{POTA}) needs to be replaced with the \emph{average} optimal social welfare $\frac{1}{T}\sum_{t=1}^{T}\textrm{OPT}^t$.

An online learning algorithm is said to successfully track the sequence of (mixed) NE in a dynamic game if the average distance between the sequence of (mixed) action profiles resulting from the algorithm and the sequence of (mixed) NE vanishes as $T$ tends to infinity. A representative study in \cite{duvocelle2018learning} considers a game with a continuum of actions and dynamicity manifesting only in the utility functions. Under the assumptions that the sequence of NE is slow-varying and the utility functions are monotonic, it was shown that learning algorithms with sublinear dynamic regret successfully track the sequence of NE. The monotonicity of the utility functions plays a key role in the analysis: it translates the closeness between the learning dynamics and the NE in terms of the cumulative reward (as in the regret measure) to the closeness in terms of their distance in the action space (the concern of the tracking~outcome).

The performance of no-regret learning in terms of social welfare was studied in \cite{lykouris2016learning} for games with a dynamic population of players. Specifically, in each stage, each player may independently exit with a fixed probability and is subsequently replaced with a new player with a potentially different utility function (the population size is therefore fixed and the player set is a stationary process over time). For structural games such as first-price auctions, bandwidth allocation, and congestion games, the relation between no-adaptive-regret learning and the average optimal social welfare was examined.

Game dynamics can be in diverse forms. There lacks a holistic understanding on the matching between regret notions and the underlying dynamics of the game. 
Different forms of game dynamics demand different benchmark policies in order to arrive at a meaningful regret measure that lends significance to the stamp of ``no-regret learning'' yet at the same time is attainable. Viewing from a different angle, one may pose the fundamental question on what kinds of game dynamics are tamable through distributed online learning and make no-regret learning and approximately optimal social welfare feasible.

\section{Incomplete and Imperfect Feedback}\label{IS2}

Learning and adaptation rely on feedback. Quality of the feedback in terms of completeness and accuracy thus has significant implications in no-regret learning. We explore this issue in this section.

\subsection{Incomplete Feedback}

Incomplete feedback stands in contrast to full-information feedback where utilities of all actions a player could have taken are observed in each stage. Incompleteness can be spatial across the action space or temporal across decision stages. In the former case, a commonly studied model is the so-called \emph{bandit feedback}, where only the utility of the chosen action is revealed. In the latter, the feedback model is referred to as \emph{lossy feedback} where there are decision stages with no feedback \cite{zhou2018learning}. One can easily envision a more general model compounding bandit feedback with lossy feedback. Studies on this general model are lacking in the literature.

The term ``bandit feedback'' has its roots in the classical problem of \emph{multi-armed bandit}~\cite{zhao2019multi}. The name of the problem comes from likening an archetypical single-player online learning problem to playing a multi-armed slot machine (known as a bandit for its ability of emptying the player's pocket). Each arm, when pulled, generates rewards according to an unknown stochastic model or in an adversarial fashion. Only the reward of the chosen arm is revealed after each play. Due to the incomplete feedback, the player faces the tradeoff between exploration (to gather information from less explored arms) and exploitation  (to maximize immediate reward by favoring arms with a good reward history).

In a multi-player game setting with bandit feedback, no-regret learning from an individual player's perspective can be cast as a single-player \emph{non-stochastic}/\emph{adversarial} bandit model where the payoff of each arm/action is adversarially chosen and aggregates the interaction with the other players in the game. The concept of external regret in the game setting corresponds to the weak regret in the adversarial bandit model \cite{auer2002nonstochastic}, which adopts the best single-arm policy in hindsight as the benchmark. The MW algorithm was modified in \cite{auer2002nonstochastic} to handle the change of the feedback model from full-information to bandit. Specifically, the weight~$W_a(t)$ of action $a$ at time $t$ is updated as $W_{a}(t)=W_{a}(t-1)e^{\epsilon r_{a}(t)/p_{a}(t)}$ where $p_{a}(t)$ is the probability of selecting action $a$ at time $t$ and $r_a(t)=0$ if $a$ is unselected. Dividing the observed reward by the corresponding probability of the chosen action ensures the unbiasedness of the observation. Quite intuitively, the price for not observing the rewards of all actions is the degradation of the regret order in the size of the action space, i.e.,~from $\Theta(\sqrt{\log(|\mathcal{A}|)T})$ in the full-information setting \cite{cesa2006prediction}, to $\Theta(\sqrt{|\mathcal{A}|T})$ in the bandit setting \cite{audibert2009minimax}. 

The multi-player bandit problem explicitly models the existence of $N$ players competing for $M$ ($M>N$) arms \cite{liu2010distributed}. Originally motivated by applications in wireless communication networks where distributed users compete for access to multiple channels, this specific game model is characterized by a special form of interaction among players: a collision occurs when multiple players select the same arm, which results in utility loss. The objective of this distributed learning problem is to minimize the \emph{system-level regret} over all players against the optimal centralized (hence collision-free) allocation of the players to the best set of arms~\cite{liu2010distributed}. In addition to the exploration-exploitation tradeoff in the single-player setting, this distributed learning problem under a system-level objective also faces the tradeoff between selecting a good arm and avoiding colliding with competing players. A number of distributed learning algorithms have been developed to achieve a sublinear system-level regret with respect to $T$ \cite{liu2010distributed}. Recent extensions of the multi-player bandit problem further consider the setting where each arm offers different payoffs across players \cite{bistritz2018distributed}.

The multi-player bandit problem is a special game model in that the players have identical action space and their interaction is only in the form of ``collisions'' when choosing the same action. In a general game setting, the impact of incomplete feedback on no-regret learning and system-level performance is largely open. One quantitative measure of the impact is the regret order with respect to the size of the action space. As mentioned above, bandit feedback results in an additional $\sqrt{|\mathcal{A}|}$ term in the regret order, which can be significant when the action space is large. Recent work \cite{cesa2019delay,bar2019individual} has shown that local communications among neighboring players in a network setting can mitigate the negative impact of bandit feedback on the regret order in $|\mathcal{A}|$. In terms of the impact on the system-level performance, it has been shown under a game model with a continuum of actions that bandit feedback degrades the convergence rate of the learning dynamics to equilibria~\cite{bravo2018bandit}.

\subsection{Imperfect Feedback}

Imperfect feedback refers to the inaccuracy of the observed utilities in revealing the quality of the selected actions. Recall that mixed strategies are necessary for achieving no-regret learning in the adversarial setting. The quality of a mixed strategy is characterized by the expected utility where the expectation is taken over the randomness of strategies of all players. Referred to as \emph{expected feedback}, the feedback model assuming observations on the expected utility, however, can be unrealistic. A more commonly adopted feedback model is the \emph{realized feedback} where only the utility of the realized action profile is revealed. The realized feedback can be viewed as a noisy unbiased estimate of the expected feedback where the noise is due to the randomness of players' strategies.

The so-called  \emph{noisy feedback} assumes a different source of noise: it comes from the external environment and is additive to either the observed utility vectors in the so-called semi-bandit feedback \cite{heliou2017learning} with a finite action space, or the gradient of the utility functions in the first-order feedback~\cite{mertikopoulos2019learning} with a continuum of actions. Under the assumptions of unbiasedness and bounded variance, the issue of the additive noise can be addressed by rather standard estimation techniques and analysis.  A more challenging setting is to consider non-stochastic noise due to adversarial attacks, especially in applications such as adversarial machine learning. This problem was recently studied in the single-player setting~\cite{jun2018adversarial}. Studies in the multi-agent setting are still lacking.

\section{Bounded Rationality}\label{IS3}

The concept of \emph{bounded rationality} was first introduced in economics \cite{simon1955behavioral} to provide more realistic models than the often adopted perfect rationality that assumes the decision-making of players is the result of a full optimization of their utilities. In reality, players often take reasoning shortcuts that may lead to suboptimal decisions. Such reasoning shortcuts may be a result of limited cognition of human minds or necessitated by the available computation time and power relative to the complexity of action optimization.

Cognitive limitations include the limited ability in anticipating other decision-makers' strategic responses and certain psychological factors that interfere with the valuation of options. Various models exist for capturing the limitations in the players' valuation of options. For example, a player may be myopic, focusing only on the short-term reward~\cite{gabaix2005bounded}. Even with forward-thinking, a player may settle for suboptimal actions perceived as acceptable by the player \cite{simon1955behavioral}. The limitation in a player's ability to anticipate other players' strategies can be modeled through a \emph{cognitive hierarchy} by grouping players according to their cognitive abilities and characterizing them in an iterative fashion. Specifically, players with the lowest level of cognitive ability are grouped as the level-$0$ players who make decisions randomly. Level-$k$ ($k>0$) players are then defined iteratively as those who assume they are playing against lower-level players and anticipate the opponents' strategies accordingly. Recent work draws an interesting connection between the cognitive hierarchy model and the \emph{Optimistic Mirror Descent} (OMD) algorithm for solving the saddle point problem with applications in generative adversarial networks \cite{daskalakis2018training}. The saddle-point problem can be viewed as a two-player zero-sum game with a continuum of actions. The solutions to the problem correspond to the set of NE. It has been shown that the OMD algorithm guarantees a converging system dynamic to an NE in terms of the actual plays while Gradient Descent~(GD) may lead to cycles \cite{daskalakis2018training}. In the language of cognitive hierarchy, players adopting GD can be regarded as level-0 thinkers in the sense that they do not anticipate the strategies of their opponents. Players adopting OMD are level-1 thinkers since they take advantage of the fact that their opponents are taking similar gradient methods, which will not lead to abrupt gradient changes between two consecutive stages \cite{daskalakis2018training}. Consequently, an extra gradient update is applied in OMD to accelerate learning.

Besides cognitive limitations, players are also constrained in terms of physical resources such as memory and computation power. Acquiring, storing, and processing all relevant information for decision-making may be infeasible, especially in complex systems with a~large action space. For example, players may only choose from strategies with bounded complexity~\cite{scarsini2012repeated}, or use only recent observations in decision-making due to memory constraints~\cite{chen2017k}.

While models for bounded rationality abound in economics, political science, and other related disciplines, incorporating such models into distributed online learning is still in its infancy. A holistic understanding on the implications of bounded rationality in distributed online learning is yet to be gained. An intriguing aspect of the problem is that bounded rationality may not necessarily imply degraded performance. For example, in dynamic games, bounded memory of past experiences may have little effect since no-regret learning dictates that the distant past be forgotten (see discussions in Sec. 3).

\section{Heterogeneity}\label{IS4}

The heterogeneity of complex multi-agent systems characterizes the asymmetry across players in three aspects: the available information and knowledge about the system, available actions, and the level of adaptivity to opponents' strategies. In the example of mixed traffic in urban transportation, autonomous vehicles, while likely to have greater computation power for solving complex decision problems, may have to obey an additional set of regulations on available actions.

In adversarial machine learning, in addition to the asymmetry on the knowledge and power, the attacker and the defender may also have different levels of real-time adaptivity to the other player's strategy. Classical regret notions such as the external regret that assumes fixed actions of the other players, while applicable to \emph{oblivious} attackers, are no longer valid under \emph{adaptive} attacks. A partial solution is to adopt a new notion of \emph{policy regret} defined against an adaptive adversary who assigns reward vectors based on previous actions of the player \cite{arora2012online}. Specifically, let $u_t(\cdot;a_{1:t-1})$ denote the player's reward function determined by the adversary at time $t$, given the sequence of actions $a_{1:t-1}$ taken by the player in the past. The policy regret with reward functions $\{u_t\}_{t=1}^{T}$ is defined as
\begin{equation}
	\max_{a\in\mathcal{A}}\mathbb{E}\left[\sum_{t=1}^{T}u_t(a; \{a,...,a\})-\sum_{t=1}^{T}u_t(a_t; a_{1:t-1})\right],
\end{equation}
where $u_t(\cdot;\{a,...,a\})$ denotes the reward function determined by the adversary if the player took actions $\{a,...,a\}$ in the past.  The $m$-memory policy regret is defined by assuming that the reward function depends only on the past $m$ actions of the player.

The difference between the external regret and the policy regret may not be crucial if the adversary and the player have homogeneous objectives (e.g., mixed traffic in transportation systems). It has been shown that there exists a wide class of algorithms that can ensure no-regret learning under both regret definitions, as long as the adversary is also using such an algorithm \cite{arora2018policy}. In applications such as adversarial machine learning where the adversary may be a malicious opponent, the two notions of regret are incompatible: there exists an $m$-memory adaptive adversary that can make any action sequence of the player with sublinear regret in one notion suffer from linear regret in the other \cite{arora2018policy}. A general technique for developing no-policy-regret algorithms in the single-player setting was proposed in \cite{arora2012online}. In terms of the system-level performance, it was shown in two-player games that no-policy-regret learning guarantees convergence of the system dynamic to a new notion of equilibrium called \emph{policy equilibrium} \cite{arora2018policy}. However, the understanding of policy equilibrium is limited. In games with more than two players, even the definition of policy equilibrium is unclear.

\bibliographystyle{IEEEtran}
\bibliography{References}
\end{document}